\documentclass[a4paper,11pt]{article}
\usepackage{pos}

\title{Quarkonium and Heavy-Flavor Transport}

\author*[a]{Xiaojun Yao}

\affiliation[a]{InQubator for Quantum Simulation, Department of Physics, University of Washington,\\
Seattle, WA 98195, USA}

\emailAdd{xjyao@uw.edu}

\abstract{This is a theoretical overview of quarkonium production in relativistic heavy ion collisions given for the Hard Probes 2026 conference held at Vanderbilt University in Nashville. The talk covers cold nuclear matter effects, various phenomenological models for describing the dynamical dissociation and recombination processes in a hot quark-gluon plasma, and the open quantum system framework, highlighting a universal Lindblad equation that unifies the quantum Brownian motion and quantum optical limits, as well as a recent development in constructing a non-Markovian evolution equation for quarkonium in-medium dynamics. Recent developments in single heavy quark dynamics are also briefly discussed.
\\
\\
}

\renewcommand{\hookAfterAbstract}{%
  \par\bigskip
  \textsc{Preprint}: IQuS@UW-21-141 
}

\FullConference{13th International Conference on Hard and Electromagnetic Probes of High-Energy Nuclear Collisions (HardProbes2026)\\
21-26 June 2026\\
Nashville, Tennessee, USA\\}

\begin{document}
\maketitle

\section{Introduction}
\label{sec:intro}
Quarkonium is a bound state of a heavy quark ($Q$) and a heavy antiquark ($\bar{Q}$). In vacuum, the potential between the $Q\bar{Q}$ pair can be well described by the Cornell potential, which contains a Coulomb part $\sim1/r$ and a confining part $\sim r$. In 1986, Matsui and Satz studied the plasma screening effect on the potential and found that at sufficiently high temperature the potential is significantly suppressed so that quarkonium ``melts'' inside a hot quark-gluon plasma~\cite{Matsui:1986dk}. Since then, quarkonium has been used as a probe of deconfinement in relativistic heavy ion collisions.

Let us ask the question of whether quarkonium can exist inside the quark-gluon plasma (QGP). If we consider a deeply bound state at low temperature, the quarkonium state is approximately a small color dipole and interacts with the QGP weakly. For example, we have $rT\sim0.15$ for $\Upsilon$(1S) at $T=200$ MeV. Moreover, a recent lattice QCD calculation extracted the spectral function from the correlator of two temporal Wilson lines in Coulomb gauge in the static limit $m_Q\to\infty$, and hinted at no screening in the real part of the potential up to $T\approx350$ MeV~\cite{Bazavov:2023dci}. The same calculation also confirmed the broadening of the spectral function at finite temperature: a sharp peak in vacuum corresponding to a quarkonium state becomes a broad one in the medium. The thermal width of the spectral function for each quarkonium state can be understood as being generated by an imaginary part of the potential that grows with both temperature and distance. Physically, it means quarkonium can dissociate by dynamically absorbing an energetic gluon from the medium, which has been perturbatively calculated before~\cite{Laine:2006ns,Beraudo:2007ky,Brambilla:2008cx,Brambilla:2011sg,Brambilla:2013dpa}. If quarkonium can exist inside the QGP, recombination can also happen as the inverse process of dissociation~\cite{Thews:2000rj}, in which an unbound $Q\bar{Q}$ pair radiates out a gluon that takes away enough energy and forms a bound state. So the physical picture has evolved from static potential screening to dynamical dissociation and recombination. Phenomenologically, the time evolution of the quarkonium density in the medium can be described in terms of dissociation and recombination contributions.

\section{Cold and Hot Nuclear Matter Effects}
\label{sec:cnm_hnm}
\begin{figure}[t]
    \centering
    \includegraphics[width=1\textwidth]{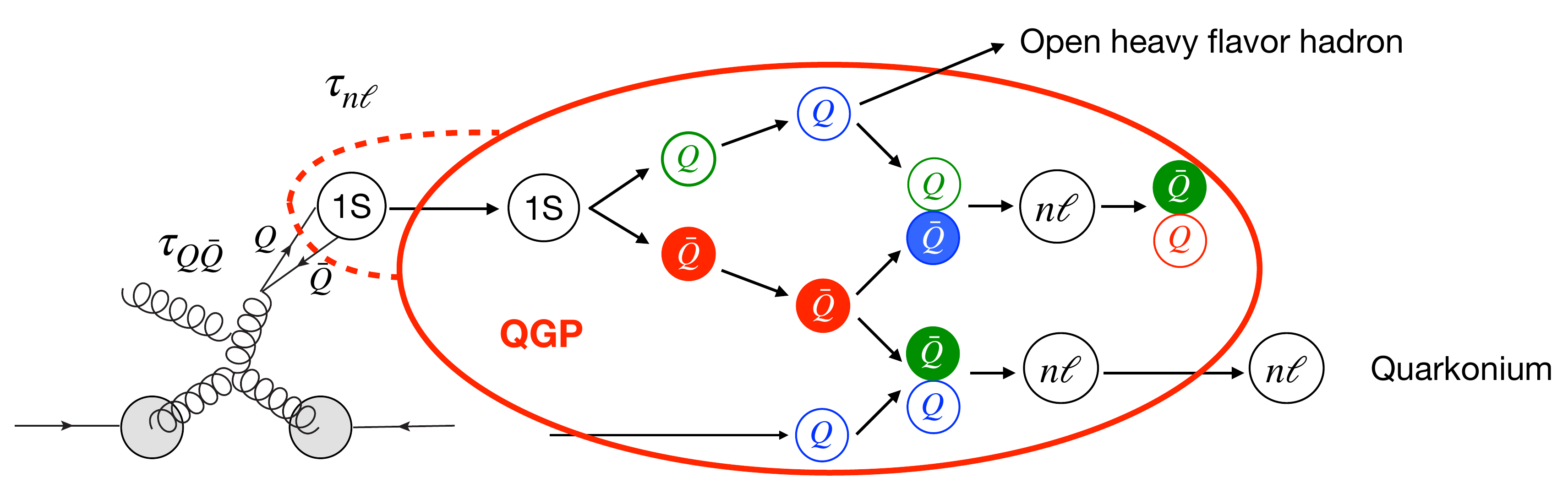}
    \caption{Schematic diagram for the physical processes relevant for quarkonium production in heavy ion collisions.}
    \label{fig:scheme}
\end{figure}

In phenomenological studies of quarkonium production in heavy ion collisions, the cold and hot nuclear matter effects have to be considered together. The relevant physical processes are schematically shown in Fig.~\ref{fig:scheme}. In the beginning, we have a $Q\bar{Q}$ pair produced from an initial hard scattering. Two gluons, one from each colliding nucleus, scatter and produce another gluon and a $Q\bar{Q}$ pair. The initial hard process is affected by the cold nuclear matter (CNM) effect on the parton distribution function (PDF) and is often modeled by nuclear PDF (nPDF). For quarkonium produced at midrapidity in high energy collisions, the relevant longitudinal momentum fraction is in the range $x\leq 0.01$, which lies in the shadowing region of the nPDF~\cite{Paukkunen:2014nqa}. Physically, shadowing means the nPDF is suppressed with respect to the PDF of a free nucleon. At lower collision energies and forward and backward rapidities, the antishadowing region may also become relevant. The cold nuclear modification factors solely induced by the nPDF for quarkonium can be calculated by using, e.g., the EPPS21 parameterization~\cite{Eskola:2021nhw}.

After the initial production of the $Q\bar{Q}$ pair, they then evolve into a quarkonium state. Other cold nuclear matter effects such as energy loss, momentum broadening, and nuclear absorption may influence the development of the quarkonium state, depending on the formation time.
Two relevant time scales matter here: the production time $\tau_{Q\bar{Q}}$ of the $Q\bar{Q}$ pair and the formation time $\tau_{n\ell}$ of a particular quarkonium bound state labeled by $n\ell$. In the rest frame of the $Q\bar{Q}$ pair, these time scales are given by $1/(2m_Q)$ and $1/\Delta E_{n\ell}$, respectively, where $\Delta E_{n\ell}>0$ denotes the binding energy of the specific quarkonium state. Due to the large heavy quark mass $m_Q$, the production of the $Q\bar{Q}$ pair is a short-distance process. On the other hand, the development of a quarkonium bound state is a long-distance process.
In the laboratory frame, we need to take into account the time dilation factor at nonzero transverse momentum $p_T$. The $Q\bar{Q}$ pair production time is roughly
\begin{equation}
\tau_{Q\bar{Q}} \sim \frac{\sqrt{4m_Q^2+p_T^2}}{2m_Q}\frac{1}{2m_Q} \,,
\end{equation}
while it takes a much longer time for the pair to develop into the specific quarkonium state
\begin{equation}
\tau_{n\ell} \sim \frac{\sqrt{4m_Q^2+p_T^2}}{2m_Q}\frac{1}{\Delta E_{n\ell}} \,.
\end{equation}
At low and intermediate transverse momenta, the $Q\bar{Q}$ pair production takes a very short time, so it is only affected by the initial-state nPDF effect, but not by any final-state nuclear medium effect. The formation time of the most deeply bound quarkonium state, $\Upsilon$(1S), is also small at low transverse momentum, probably smaller than the typical formation time of the QGP. So hot medium effects may not modify the development of $\Upsilon$(1S) from the initially produced $Q\bar{Q}$ pair. But in general, one must start accounting for the various medium effects right after the production of the $Q\bar{Q}$ pair.

Once the $Q\bar{Q}$ pair enters the QGP phase, either as an unbound pair or a bound quarkonium state, we have hot nuclear matter effects. They include dissociation and recombination mentioned earlier. If a bound quarkonium state dissociates, the two heavy quarks diffuse via interacting with the QGP, and may move away from each other. Then they hadronize into open heavy flavor hadrons individually. Or they may stay close to each other and form a bound state again later when the temperature is low and favors the formation of quarkonium states. I want to emphasize that they do not have to form the original quarkonium state from the dissociation of which they are produced. They may form any quarkonium state that is allowed to exist at a given temperature. Another scenario is that they may diffuse away from each other, meet another heavy quark that is produced from a different initial hard collision, and form a quarkonium state.
Because open heavy quarks only partially thermalize inside the QGP, it is important to track the dynamics of single heavy quarks in order to model recombination accurately. This motivated me and my collaborators to develop a coupled description of both open and hidden heavy flavors~\cite{Yao:2020xzw}.

After the hadronization of the QGP, one also needs to account for hadronic feed-down processes such as the decay of the 1P states to 1S states.

Therefore, the necessary ingredients of a phenomenological model include CNM effects, dissociation and recombination, and single heavy quark dynamics. Various models treat these ingredients differently. A community summary of these models can be found in Ref.~\cite{Andronic:2024oxz}. In the following I will discuss a few examples with a focus on new developments reported in this conference.

\section{Phenomenological Models}
\label{sec:pheno}

\subsection{Statistical Hadronization Model}
The Statistical Hadronization Model (SHM)~\cite{Andronic:2017pug,Andronic:2019wva} assumes that charm quarks are unbound inside the QGP and have reached kinetic equilibrium, which means their $p_T$ spectrum is thermal and thus soft. Charmonium is produced via an instantaneous hadronization at freezeout. The corona effect is also included, which means that if a charmonium state is produced near the edge of the QGP fireball and moves outwards, it will be treated as in p-p collisions. The model is valid at low $p_T$ and can describe the experimentally measured nuclear modification factor $R_{AA}$ of $J/\psi$~\cite{ALICE:2022jeh}. At higher $p_T$, the model underestimates the yield since the thermal spectrum is too soft. As a result, it underestimates the measured $R_{AA}$ of $\psi$(2S) because of a higher $p_T$ cut in the experimental data. The main uncertainty of the model comes from the total $Q\bar{Q}$ production cross section $\sigma_{Q\bar{Q}}$.

\subsection{Semiclassical Transport Approach}
The semiclassical transport approach tracks the time evolution of the quarkonium number or density. In the Texas A\&M University (TAMU) model~\cite{He:2021zej}, the number density $n_{\mathcal{B}}$ of a quarkonium state $\mathcal{B}$ evolves according to a rate equation (a similar equation holds for the number)
\begin{equation}
\label{eqn:rate}
\frac{{\rm d}n_{\mathcal{B}}(t,\boldsymbol{x})}{{\rm d}t} = -\Gamma\, n_{\mathcal{B}}(t,\boldsymbol{x}) + F(t,\boldsymbol{x}) \,,
\end{equation}
where $\Gamma$ is the dissociation rate and $F$ denotes the recombination contribution. In the earlier version of the model, the dissociation rate is calculated from the $2\to2$ process $g+\mathcal{B}\to Q+\bar{Q}$ and the $2\to3$ processes in the quasi-free limit, $2\times(q/g + Q\to q/g+Q)$, which means that the quarkonium state is treated as two independent heavy quarks. The recombination contribution is not directly calculated but obtained from detailed balance with a phenomenological factor $R$
\begin{equation}
F(t, {\boldsymbol x}) = R\,\Gamma\,n^{\rm eq}_{\mathcal{B}}(t, {\boldsymbol x}) \,,
\end{equation}
where $n^{\rm eq}_{\mathcal{B}}$ is the equilibrium density that is spacetime-dependent via its temperature dependence.

In the new developments~\cite{Wu:2025hlf,Wu:2025lcj,Fu:2026tke,Fu:HP2026}, the $T$-matrix summation formalism is used, in which the interaction between a heavy quark and a medium parton is resummed by using an in-medium potential. The potential can be modeled by taking inputs from lattice QCD calculations. Both the dissociation process $q/g+\mathcal{B}\to q/g+Q+\bar{Q}$ and the regeneration process $q/g+Q+\bar{Q}\to q/g+\mathcal{B}$ are calculated from the same heavy-light $T$-matrix. In other words, recombination is now calculated rather than modeled. The single heavy quark dynamics is described by a Fokker-Planck equation, where the drag and diffusion coefficients are obtained from the same interaction in the $T$-matrix approach. This gives a consistent description of both open and hidden heavy flavors.

The phenomenological results for charmonium were discussed in this conference~\cite{Fu:HP2026}. The model results for the $R_{AA}$ of $J/\psi$ and $\psi$(2S) in 5.02 TeV Pb-Pb collisions contain a primordial component, a regeneration component and the feed-down from bottom hadron decays. The regeneration component dominates at low $p_T$ in central collisions. The total results agree with the centrality and $p_T$ dependence of the experimental data fairly well. The approach has also been applied to O-O collisions, where the CNM effects are combined with the transport calculation and compared with the preliminary data on the $J/\psi$ $R_{AA}$ at both 5.36 TeV and 200 GeV~\cite{Wu:HP2026}.

\section{Cold Nuclear Matter Effects and Transport Studies in Small Collision Systems}
In order to phenomenologically study quarkonium suppression in small collision systems such as proton-nucleus and deuteron-nucleus collisions, it is crucial to understand various CNM effects. The total cold nuclear modification factor can be written as a product~\cite{Vogt:HP2026}
\begin{equation}
R_{pA}^{\rm CNM} = R_{pA}^{\rm nPDF} \times R_{pA}^{\rm eloss+broad} \times R_{pA}^{\rm abs} \,.
\end{equation}
$R_{pA}^{\rm nPDF}$ stands for the CNM effect from the nPDF. The centrality dependence of the nPDF effect is assumed proportional to the nuclear density. The energy loss and momentum broadening effects are described by a path-dependent quenching weight $R_{pA}^{\rm eloss+broad}$. Finally, the nuclear absorption $R_{pA}^{\rm abs}$ depends on how long the formed quarkonium state interacts with the nuclear matter. From fixed-target experimental studies, it is known that the absorption cross section decreases with the collision energy. A comparison with the measured nuclear modification factors of $J/\psi$ in 200 GeV d-Au collisions as a function of rapidity in different centrality classes was shown in this conference~\cite{Vogt:HP2026}. It can be seen that the nuclear absorption, a final-state effect, is crucial to explain the data in the nucleus-going direction.

With the CNM effects as a baseline, one can study the hot medium effect in small systems. The TAMU semiclassical transport approach has been applied to bottomonium production in 5.02 TeV p-Pb collisions~\cite{Thapa:2025lcf,Thapa:HP2026}, with reaction rates from perturbation theory (TAMU-P) and the nonperturbative $T$-matrix approach (TAMU-NP). The results were also compared with those of an open quantum system calculation, which I will explain later. Without regeneration, the nonperturbative rates in the TAMU-NP approach lead to too strong suppression for the excited bottomonium states $\Upsilon$(2S) and $\Upsilon$(3S) when compared with the data. After the regeneration contribution is included, the description of the data is much improved, implying that regeneration matters for bottomonium even in small collision systems.

\section{Open Quantum System Approach}
\label{sec:oqs}
In recent years, the open quantum system (OQS) framework has become a popular tool to study quarkonium in-medium dynamics~\cite{Akamatsu:2011se,Akamatsu:2014qsa}, in particular when combined with nonrelativistic QCD (NRQCD) and potential NRQCD (pNRQCD)~\cite{Brambilla:2016wgg,Brambilla:2017zei,Blaizot:2017ypk,Miura:2019ssi,Yao:2018nmy,Yao:2020eqy}. Reviews can be found in Refs.~\cite{Rothkopf:2019ipj,Akamatsu:2020ypb,Sharma:2021vvu,Yao:2021lus}. The OQS framework treats the $Q\bar{Q}$ pair and the QGP together as a closed quantum system. The $Q\bar{Q}$ pair itself is an open quantum system, whose time evolution can be obtained by tracing out the QGP degrees of freedom
\begin{equation}
\rho_{Q\bar{Q}}(t) = {\rm Tr}_{\rm QGP}\big[ U(t)\rho_{\rm tot}(0)U^\dagger(t) \big] \,.
\end{equation}
In certain limits, the evolution equation is a Lindblad equation
\begin{equation}
\label{eqn:lindblad}
\frac{{\rm d}\rho_{Q\bar{Q}}(t)}{{\rm d}t} = -i\big[H, \rho_{Q\bar{Q}}(t)\big] + \sum_{ij}D_{ij}\Big( L_i\rho_{Q\bar{Q}}(t)L_j^\dagger - \frac{1}{2}\big\{ L_j^\dagger L_i, \rho_{Q\bar{Q}}(t) \big\} \Big) \,,
\end{equation}
where $L_i$ denotes a Lindblad operator and $D_{ij}$ stands for some correlation function of the QGP, about which I will be more explicit later. The Lindblad equation contains both dissociation and recombination contributions. For a bound state, the former is given by the anticommutator term while the latter originates from the $L_i\rho_{Q\bar{Q}}(t)L_j^\dagger$ term.

\subsection{Two Simplifying Limits}
\begin{figure}[t]
    \centering
    \includegraphics[width=1\textwidth]{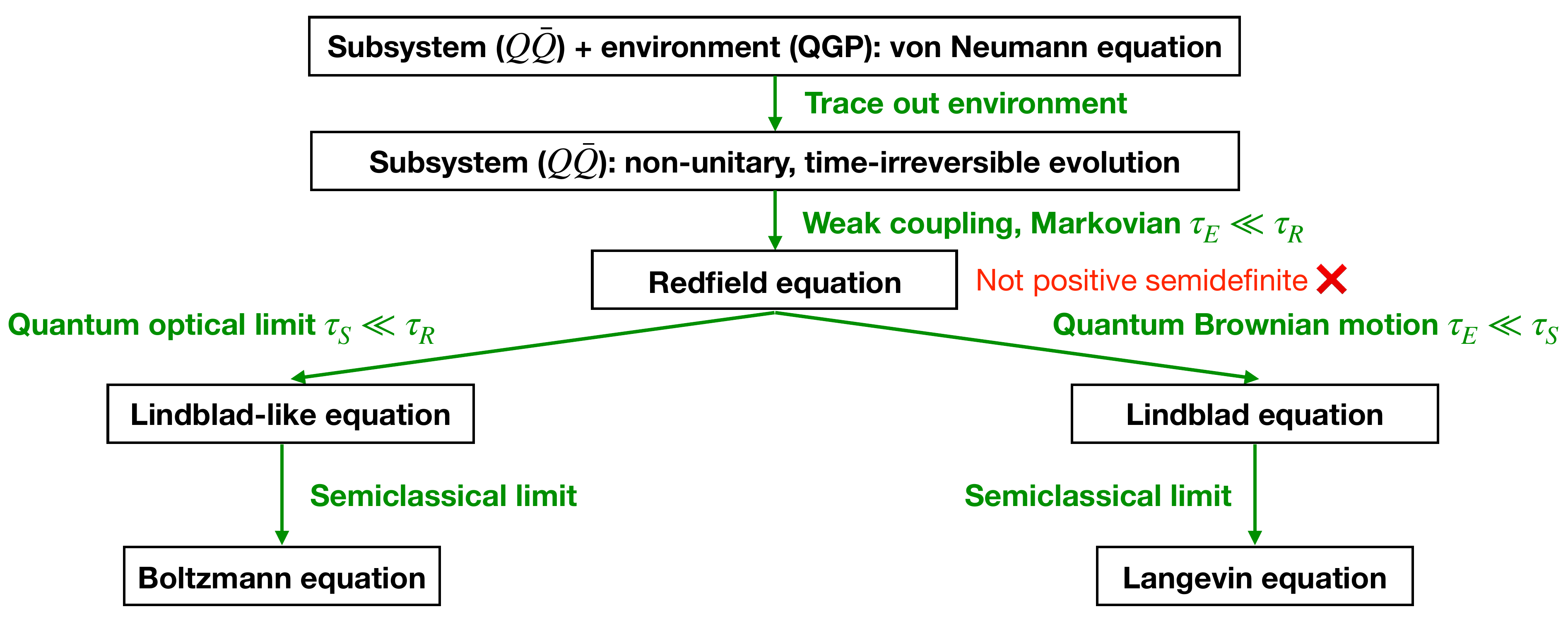}
    \caption{Sequence of approximations used in previous studies that employed the open quantum system approach for studying quarkonium in-medium dynamics.}
    \label{fig:oqs}
\end{figure}

In order to explain to you the most recent developments in using the OQS, I need to explain what has been done before. Previous studies can be summarized as in Fig.~\ref{fig:oqs}, where a sequence of approximations is illustrated. 
We start from the von Neumann equation for the total system consisting of the subsystem ($Q\bar{Q}$) and the environment (QGP), which is a unitary time evolution. After tracing out the environment, the evolution of the subsystem becomes non-unitary and time-irreversible. There are three relevant time scales here: the environment correlation time $\tau_E$, the intrinsic time scale of the subsystem $\tau_S$ and the relaxation time of the subsystem $\tau_R$. When the subsystem and the environment are weakly coupled, we have $\tau_E\ll\tau_R$, and the dynamics is Markovian. In this limit, we obtain the Redfield equation. However, the Redfield equation does not guarantee that the density matrix remains positive semidefinite. We can make one more approximation to cure this. In the quantum optical limit, $\tau_S\ll\tau_R$, we obtain a Lindblad-like equation, whose semiclassical limit corresponds to a Boltzmann equation. In the quantum Brownian motion limit, $\tau_E\ll\tau_S$, we obtain a Lindblad equation whose semiclassical limit is a Langevin equation. Physically, the quantum optical limit is valid at low temperature, where the resolving power of the QGP is weak and the dynamics is effectively described by transitions between different energy levels, e.g., 1S, 2S and unbound states. The quantum Brownian motion limit is valid at high temperature, where the resolving power of the QGP is strong and each heavy quark is jiggling around, similar to diffusion in classical Brownian motion, which leads to decoherence of the $Q\bar{Q}$ wave function. More discussions on the hierarchies of time scales and the underlying energy scales and the corresponding effective field theories can be found in Ref.~\cite{Yao:2025ivv}.

\subsection{Phenomenological Studies Using Open Quantum Systems}
The TUM-Kent approach combines the OQS in the quantum Brownian motion limit with pNRQCD under the hierarchy $m_Q\gg 1/r \gg T,\Lambda_{\rm QCD}$ and $T\gg\Delta E$~\cite{Brambilla:2020qwo,Brambilla:2023hkw,Brambilla:2025sis,Brambilla:2026evh,Brambilla:2026ivg,Brambilla:2026fxa}, where $\Delta E>0$ denotes the typical energy gap of the subsystem. The Lindblad equation is given by
\begin{equation}
\frac{{\rm d}\rho_{Q\bar{Q}}(t)}{{\rm d}t} = -i\big[H + \gamma_{\rm adj}\Delta h, \rho_{Q\bar{Q}}(t)\big] + \kappa_{\rm adj}\Big( L_{\alpha i}\rho_{Q\bar{Q}}(t)L_{\alpha i}^\dagger - \frac{1}{2}\big\{ L_{\alpha i}^\dagger L_{\alpha i}, \rho_{Q\bar{Q}}(t) \big\} \Big) \,,
\end{equation}
where $\kappa_{\rm adj}$ and $\gamma_{\rm adj}$ are two nonperturbative transport parameters, which can be extracted from the chromoelectric correlator that I will introduce later. The application of this approach to small systems was discussed in this conference~\cite{Thapa:HP2026}.

The Duke-MIT approach combines the quantum optical limit with pNRQCD under the hierarchy $m_Q\gg 1/r \gg T,\Lambda_{\rm QCD}$ and takes the semiclassical limit~\cite{Yao:2018nmy,Yao:2020eqy,Binder:2021otw}. The resulting rate equation for the density $n_{\mathcal{B}}$ of a quarkonium state $\mathcal{B}$ has the same form as Eq.~\eqref{eqn:rate} with the following dissociation rate and recombination contribution
\begin{align}
\Gamma &= \int\!\frac{{\rm d}^3\!p_{\rm rel}}{(2\pi)^3} \big|\langle \psi_{\mathcal{B}} | \boldsymbol{r} | \Psi_{\boldsymbol{p}_{\rm rel}} \rangle\big|^2 [g^{++}_{\rm adj}]^>(-\omega) \,,\\
F &= \int\!\frac{{\rm d}^3\!p_{\rm cm}}{(2\pi)^3}\frac{{\rm d}^3\!p_{\rm rel}}{(2\pi)^3} f_{Q\bar{Q}}\, \big|\langle \psi_{\mathcal{B}} | \boldsymbol{r} | \Psi_{\boldsymbol{p}_{\rm rel}} \rangle\big|^2 [g^{--}_{\rm adj}]^>(\omega) \,,
\end{align}
where $\omega=\Delta E_{\mathcal{B}} + {p_{\rm rel}^2}/{m_Q}$ and $f_{Q\bar{Q}}$ denotes a two-particle distribution function that encodes correlations between the heavy quark pair. Both terms are derived from first principles and factorized into a piece that only depends on the wave functions of the $Q\bar{Q}$ pair and a piece that only depends on the properties of the QGP. The latter is expressed in terms of a chromoelectric correlator
\begin{equation}
[g^{++}_{\rm adj}]^>(t) = \frac{g^2T_F}{3N_c} \big\langle E_i^a(t) \mathcal{W}^{ab}(t,0) E_i^b(0) \big\rangle_T \,, \quad [g^{++}_{\rm adj}]^>(\omega) = e^{\omega/T} [g^{--}_{\rm adj}]^>(-\omega) \,,
\end{equation}
where $\mathcal{W}^{ab}$ is an adjoint temporal Wilson line. $[g^{++}_{\rm adj}]^>$ differs from the correlator for heavy quark diffusion in terms of the Wilson lines~\cite{Scheihing-Hitschfeld:2022xqx}. Its zero frequency limit gives $\kappa_{\rm adj} = [g^{++}_{\rm adj}]^>(\omega=0)$. The chromoelectric correlator encodes the QGP properties relevant for quarkonium in-medium dynamics nonperturbatively and can recover previous perturbative results if it is expanded perturbatively~\cite{Nijs:2023dbc}. A Euclidean lattice calculation method for this correlator has been formulated~\cite{Scheihing-Hitschfeld:2023tuz} and studied numerically~\cite{Brambilla:2025cqy}. It has also been calculated for strongly coupled plasmas~\cite{Nijs:2023dks}.

\subsection{Universal Lindblad Equation Unifying Two Limits}
The quantum Brownian motion and quantum optical limits are not exclusive to each other. In the regime $\tau_R \gg \tau_S \gg \tau_E$, both are valid. As the QGP expands and cools down, the environment goes from the high temperature regime where the quantum Brownian motion limit is a more effective description to the low temperature regime where the quantum optical limit is more effective.
So it will be useful to construct a unified description throughout the whole time evolution.

Starting from the Redfield equation, one can derive a universal Lindblad equation~\cite{Nathan:2020ule} that works at both low and high temperatures. The key step is to write the environment correlator as a convolution
\begin{equation}
\Delta^>(t'-t'',x'-x'') = \int_{-\infty}^{+\infty}\!{\rm d}v \int_{-\infty}^{+\infty}\! {\rm d}y\, g(t'-v,x'-y)\,g(v-t'',y-x'') \,,
\end{equation}
where the function $g$ can be obtained as the square root of $\Delta^>$ in the frequency-momentum space. The resulting equation is of the Lindblad form and thus preserves the positivity of the density matrix, without assuming the quantum optical or the quantum Brownian motion limit. The universal Lindblad equation has recently been constructed for quarkonium in the QGP~\cite{DaddiHammou:2026gle} and was presented in this conference~\cite{Gossiaux:HP2026}.

\subsection{Beyond Markovian Dynamics}
Both the Redfield equation and the universal Lindblad equation are Markovian and rely on the assumption of weak coupling between the $Q\bar{Q}$ and the QGP. In a strongly coupled plasma the Markovian description of quarkonium in-medium dynamics becomes trivial, which motivates a non-Markovian description~\cite{Nijs:2023dks,Nijs:2023dbc}. Starting from the general non-unitary and time-irreversible time evolution of the subsystem shown in Fig.~\ref{fig:oqs}, one can construct a non-Markovian evolution equation for quarkonium. Phenomenological results for the $R_{AA}$ of $\Upsilon$(1S), $\Upsilon$(2S) and $\Upsilon$(3S) as functions of centrality and $p_T$ at both 5.02 TeV and 200 GeV were presented in this conference~\cite{Scheihing-Hitschfeld:HP2026poster}, which reasonably describe the main features of the experimental data.

\section{Summary}
\label{sec:summary}
In this talk, I reviewed recent developments in using quarkonium as a probe of the QGP. The physical picture has evolved from potential screening to dynamical dissociation and recombination. In the semiclassical transport approach, the $T$-matrix formalism with lattice inputs has been used and recombination is now explicitly calculated. In the open quantum system approach, a universal Lindblad equation has been constructed that unifies the quantum Brownian motion and quantum optical limits. Moreover, a non-Markovian evolution equation has been developed and applied to study bottomonium suppression. The single heavy quark dynamics also matters for quarkonium dynamics, for which I refer you to other talks in this conference~\cite{Rapp:HP2026,DuPlessis:HP2026,Krishna:HP2026}, in particular the new development on non-Gaussianity~\cite{Rajagopal:2025ukd,Rajagopal:2026sdh,Scheihing-Hitschfeld:HP2026,DuPlessis:2026ngl}. I also discussed systematic studies of CNM effects, particularly in small collision systems. Finally, we have talks covering studies of exotic heavy hadrons, which I did not have time to cover and refer you to Refs.~\cite{Liao:HP2026,Yun:HP2026}.

\vspace{1em}
This work is supported by the U.S. Department of Energy, Office of Science, Office of Nuclear Physics, InQubator for Quantum Simulation (IQuS) (https://iqus.uw.edu) under Award Number DOE (NP) Award DE-SC0020970 via the program on Quantum Horizons: QIS Research and Innovation for Nuclear Science.

\end{document}